\documentstyle[editedvolume,epsfig]{crckapb} 

\begin{opening}
\title{BEYOND IMPLIED VOLATILITY:\protect\\}
\subtitle{Extracting information from options prices}

\author{Rama CONT}

\institute{Institut de Physique Th\'eorique\\
Ecole Polytechnique F\'ed\'erale de Lausanne\\
EPFL-IPT, CH-1015 Lausanne, Switzerland\footnote{Lecture presented
at  E\"otv\"os University, 
Budapest (July 1997). To appear in: Kertesz, J. \& Kondor, I. (Eds.) {\it Econophysics: an emerging science}, Dordrecht: Kluwer. E-mail: cont@ens.fr}.}

\end{opening}

\runningtitle{BEYOND IMPLIED VOLATILITY}

\begin{document}


\begin{abstract}
After a brief review of option pricing theory, 
we introduce various methods proposed for extracting the statistical information
implicit in options prices. We discuss the advantages and drawbacks of
each method, the  interpretation of their results in economic
terms, their theoretical consequences and their relevance for applications.
\end{abstract}

\section{Introduction}

Option pricing  has become an important field of theoretical and applied 
research both in probability theory and finance, focusing the attention of many mathematicians, financial
economists and physicists.
 Meanwhile, the rapid expansion of the options market has made option
pricing and
hedging an important issue for market practitioners.
The interest for option pricing theory culminated in the attribution  of
the 1997 Nobel prize in Economics to Myron Scholes  and Robert Merton for 
their pioneering work on this subject.

While option pricing theory has traditionally focused on 
obtaining methods for pricing and hedging of derivative securities
based on parameters of the underlying assets, recent approaches
tend to consider  the market prices of options as given and 
view them as a source of information on the market. 
While there are many excellent 
textbooks and monographs on the former approach
\cite{duffie,musiela}, the latter has only been developed in the recent literature
and is less well known.
It is this approach on which we will focus here:
we will try to show why market prices of options can be considered as a
source of information, describe different theoretical tools and
procedures for
extracting their information content and show how this information can be
interpreted in economic terms and used in applications.

The following text  is divided into four sections.
Section \ref{general} is a general review of option pricing theory
and introduces notations used throughout the text.
Section \ref{inverse} discusses the informational content of option prices
and defines the notions of implied volatility and state price density.
Section \ref{methods}
presents various methods which have been 
proposed to extract the information
content of option prices.
Section \ref{results} discusses how these results may be interpreted in economic
terms
and used in applications.
Section \ref{summary}
highlights the salient features of the results obtained in various empirical
studies and the important points to keep in mind when interpreting and using them.

\section{Option pricing: a review} \label{general}

\subsection{Options and derivative securities}
A {\it derivative security} or {\it contingent claim} is a  financial asset whose (future)
payoff is defined to be a function of the future price(s) of another (or several
other) assets, called the {\it underlying assets}.
Option pricing theory  focuses on the problem of pricing and
hedging derivative securities in a consistent way given
a market in which  the underlying assets  are represented as stochastic processes.

Consider an investor participating in a stock market, where stock prices fluctuate
according to a random process.
One of the simplest types of derivative securities
is a contract which entitles its bearer to buy, 
if she wishes, one share of stock at a specified date $T$ 
in the future  for a  price $K$ specified in advance. 
Such a contract is called a {\it European call option} on the stock, 
with exercise price or {\it strike} $K$ and maturity $T$.
The stock is said to be the {\it underlying asset}. 
Let $S_t$ be the price of the underlying asset at time $t$.
If at the expiration date $T$ of the option  the stock price 
is below the exercise price i.e. $S_T \leq K$
the holder  will not exercise his option to buy :
the option will  then be worthless. 
If the stock price at expiration is above the exercise price i.e. $S_T \geq K$ then bearer can {\it exercise} the option i.e. use it to buy one share of stock at the strike price $K$  and sell it at the current price $S_T$, making a profit of $S_T - K$.  A European call option is thus equivalent to a ticket entitling the bearer to a payment of $max (0, S_T - K)$ at the expiration date $T$ of the option. The function $h(S_T) = max (0, S_T - K)$ 
is called the {\it payoff} of the option. 

A European call option has therefore a non-negative payoff in all cases:
the stock price may rise or fall but in either case the bearer
of the option will not lose money. An option may be viewed as an 
insurance against the rise of the stock price above a specified level $K$
which is precisely the exercise price. Like any insurance contract, 
an option must therefore have a certain value. 
Options are financial assets themselves and may be bought or sold in a market, 
like stocks. 
Since 1975, when the first options exchange floor was opened in Chicago, 
options have been traded in organized markets. 
The question for the buyer or the seller of an option is then: 
what is the value of such a contract? 
How much should an investor be willing to a pay for an option? 
A related question is: once an option has been sold, what strategy 
should the seller (underwriter)
of the option follow in order to minimize his/her risk of having to pay off
a large sum in the case the option is exercised? The first two questions
are concerned with {\it pricing} while the last one is concerned with {\it hedging}.
The response to these questions has stimulated a vast literature, 
initiated by the seminal work of Black and Scholes \cite{bs}, and has led to the development of a sophisticated 
theoretical framework known as {\it option valuation theory}\footnote{For a general introduction
to option markets, see \cite{cox}. A mathematical treatment is given in
\cite{duffie} or \cite{musiela}.}. 

There are a great variety of derivative securities with more complicated payoff
structures. The payoff $h$ 
may depend in a complicated fashion not only on the final
price of the underlying asset but also on its trajectory (path-dependent options).
The option may also have early exercise features (American options) or depend
on the prices of more than one underlying asset (spread options).
We will consider here only the simplest type of option, namely the European
call option defined above. In fact, contrarily to what 
is suggested by many popular textbooks, even the pricing and hedging
of such a simple option is non-trivial under realistic assumptions
for the price process of the underlying asset.

\subsection{Expectation pricing and arbitrage pricing}\label{expect}
A naive approach to the pricing of an option would be to state that
the present value of an uncertain future cash flow $h(S_T)$ is
 simple equal to the discounted  expected value of the cash flow:
\begin{eqnarray} \label{expec}
E(h) &=& e^{-r(T-t)}\int_0^{\infty} dS_T\  h(S_T) p(S_T)
\end{eqnarray}
where $p$ is the probability density function of the 
random variable $S_T$ representing  the stock price at a future date $T$.
The exponential is a discounting factor taking into account the effect
of a constant interest rate $r$. 
Under some stationarity hypothesis on the increments of the 
price process, the density $p$ may be obtained by an appropriate statistical analysis of the historical evolution of prices. For this reason we will
allude to it as the {\it historical} density. We will refer to such a pricing rule as ``expectation pricing".

However, nothing guarantees that such a pricing rule is {\it consistent}
in the sense that one cannot find a riskless strategy for making a profit
by trading at these prices. Such a strategy is called an arbitrage opportunity.
The consistency of prices requires that if two dynamics trading strategies
have the same final payoff (with probability one) then they must have the
same initial cost otherwise this will create an arbitrage opportunity
for any investor aware of this inconsistency.
This is precisely the cornerstone of the mathematical approach to
option pricing, which postulates
that in a liquid market there should be no arbitrage opportunities: the market
is efficient enough to make price inconsistencies disappear almost as soon
as they appear.

The first example of this approach was given by Black \& Scholes \cite{bs}
who remarked that when the price of the underlying asset $S_t$
is described by a geometric Brownian motion process:
\begin{eqnarray} \label{bs1}
S_t &=& \exp (\mu t + \sigma B_t)
\end{eqnarray}
where $B_t$ is a Brownian motion (Wiener) process, 
then the expectation pricing rule gives inconsistent prices:
pricing European call options according to Eq. (1) can create arbitrage
opportunities. Furthermore they showed that
requiring
the absence of arbitrage
opportunities is sufficient to define a {\it unique} price for a European call option,
independently  of the preferences of market agents. This price is given by
the Black-Scholes formula:
\begin{eqnarray}\label{blackscholes}
C_{BS}(S_t,K,\sigma,t,T) &= & S_t N(d_1) - K e^{-r(T-t)} N(d_2)\\
d_1 &=&  \frac{\ln (S_t/K) + (T-t) (r + \frac{\sigma^2}{2})}{\sigma \sqrt{T-t}} \\
d_2 &=& \frac{\ln (S_t/K) + (T-t) (r - \frac{\sigma^2}{2})}{\sigma \sqrt{T-t}}
\end{eqnarray}
where $N$ is the cumulative distribution function of a standard Gaussian
random variable:
\begin{eqnarray}
N(u) &=& \frac{1}{\sqrt{2\pi}} \int_{-\infty}^{u} \exp(-\frac{u^2}{2}) du
\end{eqnarray}
However the method initially used by Black \& Scholes \cite{bs}
 and Merton \cite{merton3} relies in an essential way on the hypothesis that
the underlying asset follows geometric Brownian motion (Eq.\ref{bs1}), which
does not adequately describe the real dynamics of asset prices.

The methodology of Black \& Scholes was subsequently generalized 
\cite{merton3,harrison1,harrison2} to diffusion
processes defined as solutions of stochastic differential equations
\begin{eqnarray}
dS_t&=&S_t (\mu(S_t) dt + \sigma(S_t) dW_t)
\end{eqnarray}
where $dW_t$ is Gaussian white noise (increment of a Wiener process)
and $\mu, \sigma$ deterministic functions of the price $S_t$.
A good introduction to arbitrage pricing techniques is given in \cite{baxter}.

\subsection{One asset, two distributions}

Even though naive, the representation Eq.\ref{expec} of the price of
an option as its expected future payoff is appealing to economic intuition:
the present value of an uncertain cash flow should be somehow related to its
expected value. Harrison \& Kreps \cite{harrison1}  have shown that that even in
the arbitrage pricing framework it is still possible express prices of contingent claims
as expectations of their payoff, but at a certain price (!): these expectations
are no longer calculated with the density $p$ of the underlying asset but
with another density $q$, different from $p$.

More precisely,  
Harrison \& Pliska \cite{harrison2} show that 
in a market where asset prices are described by stochastic processes verifying
certain regularity conditions, the absence of arbitrage opportunities
is equivalent to the existence of  a probability measure $Q$ equivalent \footnote{Two probability measures 
$Q$ and $P$ are said to be equivalent if for any event $A$, $P(A)=0$ iff $Q(A)=0$ i.e. if they define the same set of impossible events. In the case of a single asset considered here this is a rather mild restriction. 
}
 to 
$P$,
 called an {\it equivalent 
martingale measure} , such
that all (discounted) asset prices  are $Q$-martingales: that is, if one denotes by $q_{t,T}$ the conditional density of the stock
price at maturity $S_T$ under the measure $Q$ given the past history up to time $t$, then the price $X_t$ of any derivative asset with payoff $h_X$ 
verifies\footnote{The mathematical definition of a martingale also requires the finiteness of the first absolute moment $<|X_t|>$ which does not give any additional information here.}
\begin{eqnarray} \label{martingale}
e^{-r(T-t)} \int q_{t,T}(S_T) h_X(S_T) dS_T &=& X_t
\end{eqnarray}
In particular the (discounted) stock price itself is a $Q$-martingale:
\begin{eqnarray}
e^{-r(T-t)} \int q_{t,T}(S_T) S_T dS_T &=& S_t 
\end{eqnarray}
This does not imply that real asset prices are martingales or even driftless
processes: in fact there is a positive drift in most asset prices
and also some degree of predictability. Eq.(\ref{martingale})  should
be considered as a property defining $q_{t,T}$ and not as a property
of the price process $S_t$ whose probabilistic properties are
related to the historical density $p$. The density $q_{t,T}$ is merely
a mathematical intermediary expressing the relation between the
prices of different
options with the same maturity $T$. It should not be confused with the 
historical density $p_{t,T}$.

The martingale property (Eq. \ref{martingale}) then implies that the price of any
European option can be calculated as the expectation of its payoff 
under the probability measure $Q$. 
In particular then the price of any call option
is therefore given by: 
\begin{eqnarray} \label{SPD} 
C(S_t,K,T-t,r) &=& e^{-r(T-t)} \int_{0}^{\infty} max(S_T-K,0) q_{t,T}(S_T) \quad dS_T
\end{eqnarray}
Under the assumption of stationarity, $q_{t,T}$ will only depend on $\tau = T-t$
but this assumption does not necessarily hold in real markets.

The density $q$ has been given several names in the literature: ``risk-neutral probability",
``state price deflator" \cite{duffie}, state price density, equivalent
martingale measure.
While these different notions coincide in the case of the Black-Scholes model,
they correspond to different objects in the general case of an incomplete
market (see below). The term ``risk-neutral density" refers precisely
to the case where, as in the Black-Scholes model, all contingent payoffs cannot be replicated by
a self-financing portfolio strategy.
 This is not true in general, neither theoretically
nor empirically \cite{bouchaud,follmer2} so we will refrain from using the term ``risk-neutral" density. The term ``martingale measure" refers to the
property that asset prices are expected to be $Q$-martingales: again,
this property does not define $Q$ uniquely in the case of an incomplete
market. We will use the term state price density to refer to a
density $q$ such that the market prices of options can be expressed 
by Eq.(\ref{SPD}): the state price density should not be viewed as
a mathematical property of the underlying assets stochastic process
but as a way of characterizing the prices of {\it options} on this asset.

From the point of view of economic theory, one can consider the formalism introduced by Harrison \& Pliska
as an extension of  Arrow-Debreu theory \cite{debreu}
to a continuous time / continuous
state space
framework.  The state price density $q$ is thus the continuum equivalent
of the Arrow-Debreu state prices.
However, while the emphasis of Arrow-Debreu theory is on
the notion of value, the emphasis of \cite{harrison1,harrison2}
 is on the notions of dynamic hedging and arbitrage, which
are important concerns for market operators.

The situation can thus be summarized as follows. In the framework of an
arbitrage-free market,  each asset is characterized by
two different probability densities: the historical density $p_{t,T}$
which describes the random variations of the asset price $S$
between
$t$ and $T$ and the state price density $q_{t,T}$ which is used for
pricing options on the underlying asset $S$. These two densities are different
a priori and, except in very special cases such as the Black-Scholes
model \cite{bs} arbitrage arguments do not enable us to calculate
one of them given the other.

\subsection{Incomplete markets and the market measure}

The main results of
the arbitrage approach are existence theorems which state 
that the absence of arbitrage opportunities 
leads to the existence of a density $q$
such that all option prices are expectation of their payoffs with respect
to $q$ but do not say anything about the uniqueness of such
a measure $q$. Indeed, except in very special cases like the Black-Scholes
or the binomial tree model \cite{cox}
where $q$ is determined uniquely by arbitrage
conditions there are in general infinitely many densities  $q_{t,T}$
which satisfy no-arbitrage requirements. In this case the market is said
to be {\it incomplete}.

One could argue however that market prices are not unique either: there are
always two prices- a bid price and an ask price- quoted for each option.
This has led to theoretical efforts to express the bid and ask prices
as the supremum/ infimum of arbitrage-free prices, the supremum/infimum
being taken either over all martingale measures \cite{eberlein} or over a set
of dominating strategies. Elegant as they may seem, these approaches give disappointing results.
For example Eberlein \& Jacod \cite{eberlein} have shown that in the case
of a purely discontinuous price process taking the supremum/infimum
over all martingale measures leads to trivial bounds on the option prices which
give no information whatsoever: for a derivative asset with payoff $h(S_T)$,
arbitrage constraints impose that the price should lie in the interval
$[e^{-r(T-t)}h(e^{r(T-t)}S_t), S_t]$. For a call option $h(x)=max(x-K,0)$
and the arbitrage bounds become $[S_t-Ke^{-r(T-t)},S_t]$.
The lower bound is the price of a futures contract of exercise price
$K$: arbitrage arguments simply tell us that the price of an option
lies between the price of the underlying asset and the price of
a futures contract, a result which can be retrieved by elementary arguments \cite{cox}.
More importantly, the price interval predicted by such an approach is way
too large compared to real bid-ask spreads.

These results show that arbitrage constraints alone are not sufficient
for determining the price of a simple option such as a European call
as soon as the underlying stochastic process has a more complex behavior than
(geometric) Brownian motion, which is the case for real asset prices \cite{these}.
One therefore needs to use  constraints other than those imposed by arbitrage
in order to determine the market price of the option. 

One can represent the situation as if the market had chosen among all the possible
arbitrage-free pricing systems a particular one which  could
be represented by a particular martingale measure $Q$, the {\it market
measure}.
The situation may be compared to that encountered in the ergodic theory of
dynamical systems. For a given dynamical system there may be several
invariant measures. However, a given  trajectory of the dynamical
system will reach a stationary state described by a probability measure called the 
``physical measure" of the system \cite{ruelle}. The procedure by which
the physical measure is selected among all possible  invariant measures 
involves other physical
mechanisms is not described by the probabilistic formulation.

The first approach is to choose, among all state price densities $q$,
one which verifies a certain optimization criterion. The price of the option
is then determined by Eq. \ref{SPD} using  the SPD $q$ thus chosen.
The optimization criterion can either correspond to the minimization of
hedging risk \cite{follmer2} or to a certain trade-off between
 the cost and accuracy of hedging \cite{schal}. 
F\"ollmer \& Schweizer \cite{follmer} propose to choose among all martingales measures $q$
the one which is the closest to the historical probability $p$ in terms of relative entropy (see below). In any case the minimization
of the criterion over all martingale densities leads to the selection
of a unique density $q$ which is then assumed to be the state price density.

Another approach to option pricing in incomplete markets,
proposed by El Karoui {\it et al.}, is based
on dynamic optimization techniques: it leads to lower and upper bounds
on the price of options \cite{elk1}.

A different approach proposed by Bouchaud {\it et al.} is to abandon
arbitrage arguments and define the price of the option as
the cost of the best hedging strategy i.e. the hedging strategy which
minimizes hedging risk in a quadratic sense \cite{bouchaud}.
This approach, which is further developed in \cite{book}
is not based on arbitrage pricing
and although the prices obtained coincide with
the arbitrage-free ones in the case where arbitrage arguments define
a unique price,  they may not be arbitrage-free a priori in the mathematical
sense of the term. In particular they are not necessarily the same
as the ones obtained by the quadratic risk minimization approaches
of   F\"ollmer \& Schweizer \cite{follmer} and Sch\"al \cite{schal}.

\section{Option prices as a source of information} \label{inverse}

The options market has drastically changed since Black \& Scholes
published their famous article in 1973; today, many options are
liquid assets and their price is determined by the interplay between
market supply and demand. ``Pricing" such options may therefore
not be the priority of market operators since their market price
is an observation and not a quantity to be fixed by a mathematical
approach\footnote{Note however that hedging remains an important issue
even for liquid options.}. This has led in the recent years to
the emergence of a new direction in research: what can the observed market
prices of options tell us about the statistical properties of
the underlying asset? Or, in the terms defined above: what can 
one infer for the densities $p$ and $q$ from the observation of
market prices of options?

\subsection{Implied volatility and the smile effect}
In the Black-Scholes lognormal model, all option prices are described
by a single parameter: the volatility of the underlying asset.
Therefore the knowledge of either the price or the volatility
enables to calculate the other parameter. 
In practice, the volatility is not an observable variable whereas the
market price $P$ of the option is; one can therefore invert
the Black-Scholes formula to determine the value $\sigma_{BS}$
of the volatility parameter which would give a Black-Scholes price
corresponding to the observed market price:
\begin{eqnarray}\label{implied}
\exists \sigma_{BS}(K,T),\qquad C(S_t,K,\sigma_{BS}(K,T),T) = P 
\end{eqnarray}
This value is called the (Black-Scholes) {\it implied volatility} \cite{schmalensee}.
$\sigma_{BS}(K,T)$ can be obtained through a numerical resolution of the
above equation. 
Actually this is how the Black-Scholes
formula is used by options traders: not so much as  a pricing tool
but as a means for
switching back and forth between market prices of options and
their associated 
implied volatilities.

The implied volatility is the simplest example of a statistical parameter
implicit in option prices.
Note that the implied volatility
is not necessarily equal to the variance of the underlying asset's return:
it is extracted from option prices and not from historical data from the
underlying asset. In general the two values are different. It has been conjectured
that the implied volatility is a good predictor of the future volatility of the underlying asset
but the results highly depend on the type of data and averaging period used
to calculate the volatility \cite{chiras,schmalensee}.

\begin{figure}\label{fig1}
\epsfig{file=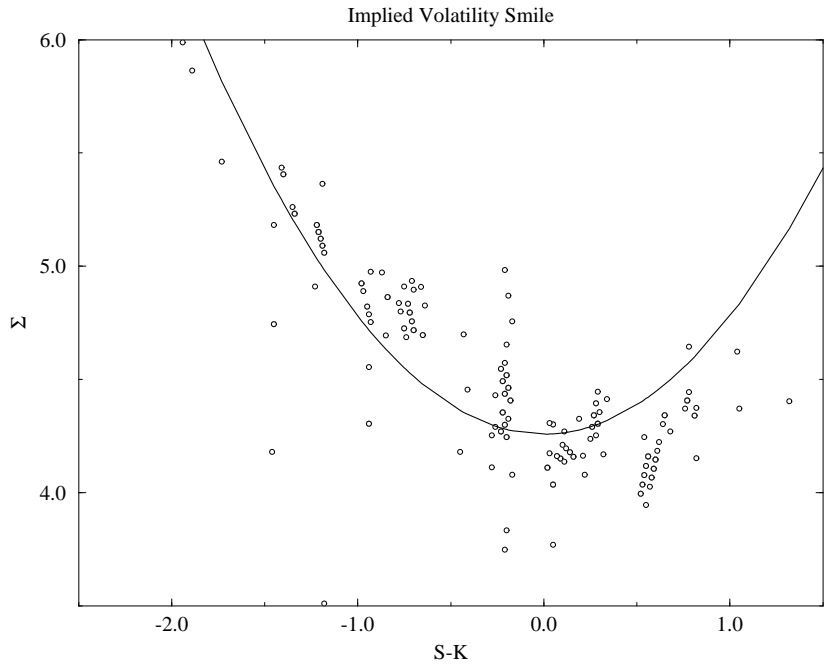,width=10cm,angle=270}
\vspace{1cm}  
\caption{Implied volatility smile for {\sc BUND} options traded on the London Financial Futures Exchange ({\sc LIFFE}). }
\end{figure}

In a  Black-Scholes universe, the implied volatility  $\sigma_{BS}(K,T)$
would in fact be a constant equal to the true volatility of the underlying asset.
The non-dependence of implied volatility on the strike price $K$ can be viewed
as a specification test for the Black-Scholes model.
However, empirical studies of implied volatilities show a systematic dependence
of implied volatilities on the exercise price and on maturity \cite{dumas,jackwerth}.
In many cases the  implied volatility presents a minimum at-the-money 
(when $S_t=K$) and has a convex, parabolic shape called the ``smile" \cite{jackwerth,adap} an example of which is given in figure \ref{fig1}.
This is not always the case however: the implied volatility plotted as a function
of the strike price $K$ may take various forms. Some of these
alternative patterns, well known to options traders, are 
documented in \cite{dumas}.  On many markets, the  convex parabolic 
``smile" pattern observed frequently after the 1987 crash has  been replaced in the recent years
by a still convex but monotonically decreasing profile.

\subsection{Implied distributions}

The empirical evidence alluded to above points out to the misspecification
of the Black-Scholes model and call for a satisfying explanation.
If the SPD is not a lognormal then there is no reason that a single parameter,
the implied volatility, should adequately summarize the information
content of option prices.
On the other hand, the availability of large data sets of option prices
from organized markets such as the CBOE (Chicago Board of Options Exchange)
add a complementary dimension to the data sets available for empirical research 
in finance: whereas time series data give one observation per date, options
prices contain a whole cross section of prices for each maturity date
and thus enable comparison between cross sectional and time series
information, giving a richer view of market variables.

In theory,  the information content of option prices is fully reflected
by the knowledge of the entire density
$q_{t,T}$: this has led to developments of methods which,  starting from
a set of option prices search for a density $q$ such that Eq.(\ref{SPD})
holds.  Such a distribution $q$ is called an {\it implied distribution},
by analogy with implied volatility\footnote{Note that except when $q_{t,T}$
is a lognormal, the variance of the implied distribution $q_{t,T}$ does
{\it not} coincide with the Black-Scholes implied volatility $\sigma_{BS}$}.
If one adheres to the assumption of absence of arbitrage opportunities,
the notion of implied distribution coincides with the concept of state price
density defined above. But even if one does not adopt this point of view,
the implied distribution still contains important information on the market.
In the following we will use indifferently the terms ``implied distribution"
and ``state price density" for $q_{t,T}$.
 Let us now
describe various methods for extracting information
about the state price density from option prices.

\section{Estimating state price densities} \label{methods}

Given that all options prices can be expressed in terms of a single function,
the state price density $q$, one can imagine statistical procedures to extract
$q$ from a sufficiently large set of option prices. Different methods have been
proposed to reach this objective, among which
we distinguish three different approaches. Expansion methods use a series expansion
of the SPD which is then truncated to give a parametric approximation, the parameters
of which can  be calibrated to observed option prices. 
Non parametric methods do not make any specific assumption on the form
of the SPD but require a lot of data.
Parametric methods postulate a particular form for the SPD and
fit the parameters to  observed option prices.

\subsection{Expansion methods}
We regroup in this section various methods which have in common the
use of a series expansion for the state price density.
The general methodology can be stated as follows. One starts with an expansion
formula for the state price density considered as a general probability distribution:
\begin{eqnarray}
P(S_T - S_t \leq x)&=& P_0(x) + \sum_{k=1}^{\infty} u_k P_k(x)
\end{eqnarray}
the first term of the expansion $P_0$ corresponding either to the lognormal or the normal
distribution. The following terms can be therefore considered as
successive corrections to the lognormal or normal approximations.
The series is then truncated  at a finite order, which gives a parametric approximation
to the SPD which, if analytically tractable, enables explicit expressions to be obtained
for prices of options.
These expressions are then used to estimate the parameters of the model from
market prices for options.
Resubstituting in the expansion enables to retrieve an approximate expression for
the SPD.

 A general feature of these methods is that
 even when the infinite sum in the expansion represents a probability distribution,
 finite order approximations of it may become negative which leads to negative probabilities
 far enough in the tails. This drawback should not be viewed as prohibitive
 however: it only means that these methods should not be used to price
 options too far from the money.

We will review here three expansion methods: lognormal Edgeworth expansions
 \cite{jarrow1}, cumulant expansions \cite{adap} and
Hermite polynomials \cite{abken}.
 
\subsubsection{Cumulants and Edgeworth expansions}
All these methods are based on a series expansion of the
Fourier transform of a probability distribution $q$ (here, the state
price density) defined by:
\begin{eqnarray}
\Phi_T(z) &=& \int q_{t,T}(x) e^{izx} dx
\end{eqnarray}
The cumulants of the probability density $p$ are then defined
as the coefficients of the Taylor expansion:
\begin{eqnarray}
\ln \Phi_T(z) &=&  \sum_{j=1}^{n} c_j(T)\frac{(iz)^j}{j!} + o(z^n)
\end{eqnarray}
The cumulants are related to the central moments $\mu_j$ by the relations
\begin{eqnarray*}
c_1 & = & \mu_1 \\
c_2 & = & \sigma^2\\
c_3 & = &  \mu_3  \\
c_4  & = & \mu_4 - 3 \mu_2 ^2
\end{eqnarray*}
One can  normalize the cumulants $c_j$ to obtain dimensionless
quantities:
\begin{eqnarray}
\lambda_j &=& \frac{c_j}{\sigma^{j}}
\end{eqnarray}
$s=\lambda_3$ is called the skewness of the distribution $p$,
$\kappa= \lambda_4$ the kurtosis. The skewness is a measure of asymmetry of
the distribution: for a distribution symmetric around its mean $s=0$, while
$s>0$ indicates more weight  on  the right side of the distribution.
The kurtosis measures the fatness of the tails: $\kappa=0$ for a normal
distribution, a positive value of $\kappa$ indicated a slowly decaying tail
while distributions with a compact support often have negative kurtosis.
A distribution with $\kappa>0$ is said to be leptokurtic.
An Edgeworth expansion  is an 
expansion of the difference between
two probability densities $p_1$ and $p_2$ in terms of their cumulants:
\begin{eqnarray}\label{edgeworth}
p_1(x) - p_2(x) &=& \frac{c_2(p_1) - c_2(p_2)}{2} \frac{d^2p_2}{dx^2} -  
 \frac{c_3(p_1) -c_3(p_2)}{3!} \frac{d^3p_2}{dx^3}\nonumber \\
& &+\frac{c_4(p_1) -c_4(p_2)+ 3(c_2(p_1) - c_2(p_2))^2 }{4!} \frac{d^4p_2}{dx^4} 
+ ...
\end{eqnarray}

\subsubsection{Lognormal Edgeworth expansions}

Since the density of reference used for evaluating payoffs in the Black-Scholes
model is the lognormal density, Jarrow \& Rudd \cite{jarrow1} suggested
the use of the expansion above, taking $p_1=q$
as the state price density and  $p_2$ as the lognormal density.
The price of a call option, expressed by Eq. \ref{SPD}, is given by:
\begin{eqnarray}\label{jarro}
C(S_t,K,T) &=&C_{BS}(S_t,K,\sigma,T) -(s-s_{LN}) e^{-rT} (\frac{\sigma^{3}}{3!} \frac{dLN(K)}{dK})\nonumber \\
& &+ (\kappa-\kappa_{LN}) e^{-rT} (\frac{\sigma^4}{4!}\frac{d^2LN(K)}{dK^2}) +...
\end{eqnarray}
where $C_{BS}$ is the Black-Scholes price, $\sigma$ the
 implied variance of the SPD and $s,s_{LN}$ and $\kappa,\kappa_{LN}$ are
respectively the skewness and the kurtosis of the SPD and of the lognormal
distribution. Given a set of option prices for maturity $T$,
 Eq. \ref{jarro}
can then be used to determine the implied variance $\sigma$
and the implied cumulants $s$ and $\kappa$.

This method has been applied by Corrado \& Su \cite{corrado}
to S\&P options: they extract the implied cumulants $s$ and $\kappa$
for various maturities from option prices and show evidence of
significant kurtosis and skewness in the implied distribution.
Using the representation above, they propose to correct the Black-Scholes
pricing formula for  skewness and kurtosis
by adding the first two terms in Eq.\ref{jarro}.
No comparison is made however between implied and historical parameters
(cumulants of $p_{t,T}$).

\subsubsection{Cumulant expansions and smile generators}
Another method, proposed by Potters, Cont \& Bouchaud \cite{adap}, is based on
 an expansion of the state price  density $q$ 
starting from a normal distribution. 
\begin{eqnarray} \label{cumu}
Q(x) - \Phi(x) &=& \frac{1}{\sqrt{2\pi}} e^{-x^2 /2} \sum_{k=3}^{\infty} P_k(x)
\end{eqnarray}
The first two terms are given by
\begin{eqnarray}
P_3(x) &=& \frac{s}{6} (1-x^2)\\
P_4(x) &=& \frac{10}{24} s^2 x^5 + (\frac{1}{24} \kappa - \frac{5}{36} s^2 ) x^3
+ (\frac{5s^2}{24} -\frac{\kappa}{8}) x
\end{eqnarray}
where $s$ is the skewness and $\kappa$ the kurtosis of the density $q$.
Although the mathematical starting point here is quite similar to 
the Hermite or Edgeworth expansion, the procedure used by Potters {\it et al}
is very different: instead of directly matching the parameters to
option prices,  they focus on reproducing correctly the shape of the volatility
smile. Their procedure is the following: starting from the expansion
(\ref{cumu}) an analytic expression for the option price can be obtained
in the form of series expansion containing the cumulants. 
The series is then truncated at a finite order $n$ and the expression for the option
price inverted to give an analytical approximation for the volatility
smile in terms of the cumulants up to order $n$, expression $\sigma(K)$
as a polynomial of degree $n$ in $K$. This expression is then
fitted to the observed volatility smile (for example using a least squares method)
to yield the implied cumulants.

An advantage of this formulation is that it corresponds more closely
to market habits: indeed, option traders do not  work with prices but with implied
volatilities which they rightly considered to be more stable in time than
option prices.

\begin{figure}\label{fig2}
\epsfig{file=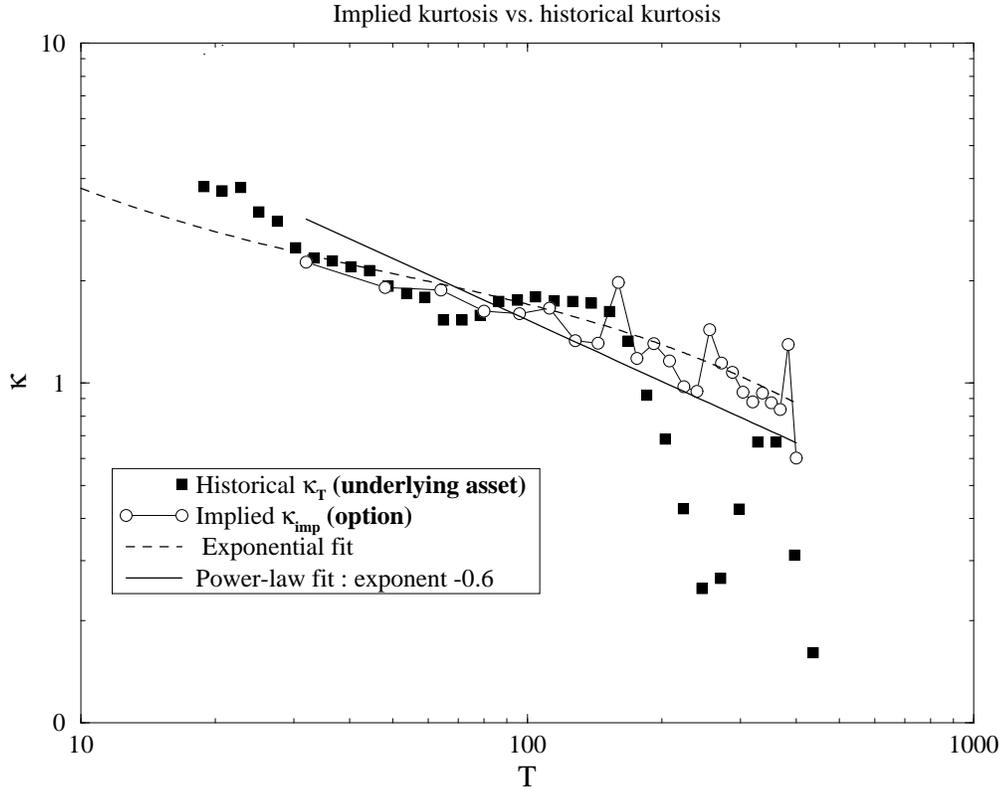,width=12cm,angle=270}
\vspace{1cm}  
\caption{Comparison of the implied kurtosis with the historical
kurtosis as a function of maturity T (in units of 5 minutes) for {\sc BUND} options traded on the London Financial Futures Exchange ({\sc LIFFE}). The representation on a logarithmic scale shows an dependence on T different from that of a random walk
for which the kurtosis would decrease as 1/T.}
\vspace{0.5cm} 
\end{figure}

This analysis can be repeated for different maturities to yield the implied cumulants
as a function of maturity $T$: the resulting term structure of the cumulants
then (shown in Fig.\ref{fig2}) gives an insight  into the evolution of the state price density $q$ 
under time-aggregation. By applying this method to options on BUND contracts
on the LIFFE market, the authors show that  the  term structure of
the implied kurtosis matches closely that of the historical
kurtosis, at least for short maturities for which kurtosis effects are
very important. This observation shows that the densities
$p_{t,T}$ and $q_{t,T}$ have similar time-aggregation properties (dependence
on $T-t$) a fact which is not easily explained in the arbitrage pricing
framework where the relation
between $p_{t,T}$ and $q_{t,T}$ is unknown in incomplete markets.

Although the expansion given in \cite{adap} uses  only the
skewness and kurtosis, one could in  principle move further in the expansion
and use higher cumulants, which would lead to a polynomial expression for the 
implied volatility smile. However  empirical estimates of
higher  order cumulants are unreliable because of their high standard deviations.

\subsubsection{Hermite polynomial expansions}
The $k$-th Hermite polynomial is defined as:
\begin{eqnarray}
\phi_k(x) &=& e^{x^2/2}\frac{d^k\phi_0}{dx^k}\qquad \phi_0(x)= \frac{e^{-x^2/2}}{\sqrt{2\pi}}
\end{eqnarray}
The method recently proposed by Abken {\it et al} \cite{abken}
uses a  Hermite polynomial expansion for both  $Q$ and the payoff function $h$.
Although the starting point is similar to the approach of \cite{adap},
the method is different: it is based on the properties of Hermite polynomials
which form an orthonormal basis for the scalar product:
\begin{eqnarray}
<f,g>= \frac{1}{\sqrt{2\pi}}\int g(x) f(x) e^{-x^2/2} dx
\end{eqnarray}
The state price density $q_{t,T}$ can be expanded on this basis:
\begin{eqnarray} \label{hermite}
q_{t,T}(x) &=& \frac{1}{\sqrt{2\pi}}\sum_{k=0}^{\infty} q_k \phi_k (x) e^{-x^2/2}\qquad 
q_k = \int q_{t,T}(x) \phi_k(x) dx 
\end{eqnarray}
Madan \& Milne also use a representation of  the payoff function $h$ 
 in the Hermite polynomial
basis:
\begin{eqnarray}
h(x) &=& \sum_{k=0}^{\infty} a_k \phi_k (x)\qquad 
a_k = <h, \phi_k>
\end{eqnarray}
Therefore, in contrast
with the cumulant expansion method,
not only the SPD is approximated but also the payoff.
The coefficients $a_k$ can be calculated analytically
for a given  payoff function $h$. In the case of a European call
the coefficients are given in \cite{abken}.
The price $C_h$ of an option with payoff $h$ is then given by:
\begin{eqnarray}\label{hermit}
C_h &=& e^{-r(T-t)}\int q_{t,T}(x) h(x) dx =
e^{-r(T-t)} \sum_{k=0}^{\infty} a_k q_k \qquad
\end{eqnarray}
The price of any option can therefore be expressed as a linear combination
of the coefficients $q_k$, which correspond to the market price
of ``Hermite polynomial risk". In order to retrieve these coefficients
from option prices, one can truncate the expansion in Eq.\ref{hermit}
at a certain order $n$ and, knowing the coefficients $a_k$, calculate
$q_k, 1 \leq k\leq n$ so as to reproduce as closely as possible a set of 
option prices, for example using a least-squares method.
The state price density can then be reconstructed using Eq. \ref{hermite}.
An empirical example is given in \cite{abken} with $n=4$: 
the empirical results show that both the historical and state price densities
have significant kurtosis; however, the tails of the state price density
are found to be fatter than those of the historical density, especially the
left tail: the interpretation is that the market fears large negative 
jumps in the price that have not (yet!) been observed in  the recent price history.
Such results have also been reported in  several other studies
of  option prices \cite{bates,yared}. 

\subsection{Non-parametric methods}
One of the drawbacks of the Black-Scholes model is that it is based on a strong
assumption for the form of the distribution of the underlying assets fluctuations,
namely their lognormality. Although everybody agrees on the weaknesses of the lognormal
model, it is not an easy task to propose a stochastic process reproducing
in a satisfying manner the dynamics of asset prices \cite{these}.

Non-parametric methods enable to  avoid this problem by using model-free 
statistical methods based on very few assumptions about the process generating
the data. Two types of non-parametric methods have been proposed
in the context of the study of option prices: kernel regression and
maximum entropy techniques.

\subsubsection{Kernel estimators}
Ait Sahalia \& Lo \cite{aitsahalia2}
have introduced another method based on the following observation 
by Breeden \& Litzenberger \cite{breeden}: if $C(S_t,K,r)$ denotes the price
of  call option then the state price density can be obtained by taking the second
derivative of $C$ with respect to the exercise price $K$:
\begin{eqnarray} \label{bibi}
q(S_T)&=& \exp^{r(T-t)} \frac{\partial^2 C}{\partial K^2}
\end{eqnarray}
If one observed  a sufficient range of exercise prices $K$
then Eq. \ref{bibi} could be used in discrete form to estimate $q$.
However it is well known that the discrete derivative of an empirically estimated curve
need not necessarily yield a good estimator of the theoretical derivative, let alone
the second-order derivative. Ait-Sahalia \& Lo propose to avoid this difficulty by
using non-parametric kernel regression   \cite{hardle}:
kernel methods yield a smooth estimator of the function $C$ and under certain
regularity conditions it can be shown that the second derivative of the estimator
converges to $q(S_T)$ for large samples. However the convergence is slowed down
both because of differentiation and because of the ``curse of dimensionality"
i.e. the large number of parameters in the function $C$.

Applying this method to S\&P futures options,
Ait-Sahalia \& Lo obtain an estimator  of the state price density for
various maturities varying between 21 days and 9 months. 
The densities obtained are systematically different from
a lognormal density and present significant skewness and kurtosis.
But the interesting feature of this approach is that it yields 
the entire distribution and not only the moments or cumulants.
One can then plot the SPD and compare it with the historical density or
with various analytical distributions.
Another important feature is that the method used by Ait-Sahalia \& Lo
also estimated the dependence on the maturity $T$ of the option prices,
yielding,
as in the cumulant expansion method, the term-structure (scaling behavior)
of various statistical parameters as a by-product.
However it is numerically intensive and difficult to use in real-time
applications.

\subsubsection{Maximum entropy method}

The non-parametric  methods described above minimize the distance
between observed option prices and theoretical option prices
obtained from a certain state price density $q$. However such a problem
has in principle an infinite number of solutions since the density $q$
has an infinite number of degrees of freedom constrained by a finite number
of option prices. Indeed, different non-parametric procedures will not lead
in general to the same estimated densities.
This leads to the need for a criterion to choose
between the numerous densities reproducing correctly the observed prices.

Two recent papers \cite{buchen,stutzer}
have proposed a method for estimating the state price density
based on a statistical mechanics/ information theoretic approach, namely
the maximum entropy method. The entropy of a probability density $p$ is defined as:
\begin{eqnarray}
S(p) &=& - \int_{0}^{\infty} p(x) \ln p(x) \quad dx
\end{eqnarray}
$S(p)$ is a measure of the information content
The idea is to choose, between all 
densities which price correctly an observed set of options, the one which
has the maximum entropy i.e. maximizes $S(q)$ under the constraint:
\begin{eqnarray}
 \int_{0}^{\infty} q(x)  \quad dx &=& 1
\end{eqnarray}
and subject to the constraint that a certain set of observed option prices $C_i$
are correctly reproduced:
\begin{eqnarray}
C_i &=& e^{-r(T-t)} \int_{0}^{\infty} max(S-K_i,0) q(S) \quad dS
\end{eqnarray}
This approach is interesting in several aspects. First, it is 
based on the minimization of an information criterion which seems
less arbitrary than other penalty functions such as those used in other non-parametric
methods.
Second, one can generalize this method to minimize the Kullback-Leibler
distance between
$q$ and the historical density $p$, defined as:
\begin{eqnarray}
S(p,q) &=& \int p(x) \ln \frac{p(x)}{q(x)} \quad dx
\end{eqnarray}
Minimizing this distance gives the state price density $q_0$ which is
the ``closest" to the historical density $p$ in an information-theoretic
sense. This density $q_0$ should be related to the minimal martingale measure
proposed by \cite{follmer}.
The value of $S(p,q)$ can then give a straightforward answer to the question
\cite{adap}: 
how different is the SPD from the historical distribution?
Or:  how different are market prices of options from those obtained
by naive expectation pricing (see section \ref{expect})?

However the absence of smoothness constraints has its drawbacks.
One of the characteristics of this method is that it typically
gives ``bumpy"
i.e. multimodal estimates of the state price density.
This is due to the fact that, contrarily to the
there is no constraint on the smoothness
of the density. This may seem a bit strange because it is not the type
of feature one expects to observe: for example, the historical PDFs of
stock returns are always unimodal. This has to contrasted with
the high degree of smoothness required in kernel regression methods.
Some authors have argued that these bumps
may be ``intrinsic properties" of market data and should not be dismissed
as aberrations but no economic explanation has been proposed.
Jackwerth \& Rubinstein \cite{rubinstein} solve this problem by
imposing smoothness constraints on the density: this can be done by
subtracting from the optimization criterion $S(q)$ a term penalizing large
variations in the derivative $dq/dx$. However the relative weight of 
smoothness vs.entropy terms may modify the results.

\subsection{Parametric methods}

\subsubsection{Implied binomial trees}

Apart from the Black-Scholes model, the other widely used option pricing
model is the discrete-time binomial tree model \cite{cox}. In the same way
that continuous-time models can be used to extract continuous state price densities
from market prices of options, the binomial tree model can be used to extract
from option prices an ``implied tree" the parameters of which are
conditioned to reproduce correctly
a set of observed option prices.
Rubinstein \cite{rubinstein94} proposes an algorithm which, starting from a set of
option prices at a given maturity, constructs an implied binomial tree
which reproduces them exactly.
The implied tree contains the same type of information as the state price density
presented above.
The tree can then be used to price other options.
Although discrete by definition, binomial trees can approximate as closely as one wishes any continuous
state price density provided the number of nodes is large enough.

Rubinstein's approach is easier to implement from a practical point of view
than kernel methods and can perfectly fit a given set of option prices for
any single maturity. However the large number of parameters may be a drawback
when it comes to parameter stability: in practice the nodes of the binomial
tree have to be recalculated every day and, as in the case of the
Black-Scholes implied volatility, the implied transition probabilities
will in general change with time.

\subsubsection{Mixtures of lognormals}
The habit of working with the lognormal 
distribution by reference to the Black-Scholes model has 
led to parametric models representing the state price density
as a mixture of lognormals of different  variances:
\begin{eqnarray}
q(S_T, T,t, S_t) &=& \sum_{k=1}^{N} \omega_k LN(\frac{S_t -\mu_k}{\sigma_k  \sqrt{T}}) \qquad\sum_{k=1}^{N} \omega_k =1
\end{eqnarray}
where $LN(x)$ is a lognormal distribution with unit variance and mean $r$, the
(risk-free) interest rate.
The advantage of such a procedure is that the price of an option is
simply obtained as the average of Black-Scholes prices for the different
volatilities $\sigma_k$ weighted by the respective weights $\omega_k$ of each distribution
in the mixture. In principle one could interpret such a mixture as
the outcome of a  switching procedure between regimes of different
volatility, the conditional SPD being lognormal in each case.
 Such models have been fitted to options prices in various markets by \cite{melick}.

Their results are not surprising: by construction, a mixture of lognormals has thin tails unless one allows high values of variance.
But the major drawback of such a parametric form  is probably 
 its absence of theoretical or economic justification.
Remember that the density which is modeled as a mixture of lognormals
is not the historical density but
the state price density: even in the hypothesis of market completeness
it is not clear what sort of stochastic process for the underlying asset
would give rise to such a state price density.

\begin{table}[htb]
\begin{center}
\caption{Advantages and drawbacks of various methods for extracting 
information from 
option prices.}
\begin{tabular}{|l|l|l|}
\hline 
Method & Advantage & Disadvantage\\
\hline
Mixture of lognormals & Link with Black-Scholes & Too thin tails \\
\hline
Expansion methods &Easy to implement and interpret& Negative tails\\
\hline
Maximum entropy& Link with historical probability& Multimodality\\
\hline
Kernel methods & Gives the entire distribution& Slow convergence\\
\hline
Implied trees & Perfect fit of cross-sectional data&Parameter instability \\
\hline
\end{tabular}
\end{center}
\end{table}

\section{Applications} \label{results}

\subsection{Measuring investors' preferences}

If one considers a simple exchange economy \cite{lucas} with a representative investor
then from the knowledge of any two of the three following ingredients 
it is theoretically possible to deduce the third one: 
\begin{enumerate}
\item The preferences of the representative investor.
\item The stochastic process of the underlying asset.
\item The prices of derivative assets.
\end{enumerate}
Therefore, at least in theory, knowing the prices of a sufficient number of options and
using  time series data to obtain information about the price process
of the underlying asset one can draw conclusions about the characteristics
of the representative agents preferences.
Such an approach has been proposed by Jackwerth \cite{jackwerth}
to extract the degree of 
risk aversion of investors implied by option prices.

Exciting as it may seem, such an approach is limited as it stands
for several reasons. First, while a representative investor approach may
be justified in a normative context (which is the one adopted implicitly
in option pricing  theory) it does not make sense in a {\it  positive}
approach to the study of market prices.
The limits of the concept of a representative agent have been already pointed out by 
many authors. Taking seriously the idea of a representative investor would
imply all sorts of paradoxes, the absence of trade not being the least of them.
Furthermore, even if the  representative agent model were qualitatively
correct, in order to obtain quantitative information on 
their preferences one must choose a parametric representation for the
decision criterion adopted by the representative investor. Typically, this
amounts to postulating that the representative investor maximizes
the expectation of a certain utility function $U(w)$ of her wealth $w$;
depending on the choice for the form of the function one may obtain different
results from the procedure described above. Given that utility functions
are not empirically observable objects, the choice of a parametric
family for $U(x)$ is often ad-hoc thus reducing the interest of such
an approach from an empirical point of view.

\subsection{Pricing illiquid options}

All options traded on a given underlying asset do not have the same
liquidity: there are typically a few strikes and maturities
for which the market activity
is intense and the further one moves away from the money
and towards longer maturities the less liquid the options become.
It is therefore reasonable to consider that some options
prices are more ``accurate" than others in the sense their prices
are more carefully arbitraged.
These considerations must be taken into account when choosing the data
to base the estimations on; for further discussion of this issue see
\cite{aitsahalia2}.

Given this fact, one can then use the information contained in the market prices
of liquid options -considered to be priced more ``efficiently"-
to price less liquid options in a coherent, arbitrage-free fashion.
The idea is simple: first, the state price density is estimated by one of the methods
explained above based only on market prices of liquid options; the estimated SPD
is then used to calculate values of other, less liquid options.
This method may be used for example to interpolate between existing 
maturities or exercice prices.

\subsection{Arbitrage strategies}

If one has an efficient method for pricing illiquid options 'better' than
the market then such a method can potentially be used for
obtaining profits by systematically buying underpriced options
and selling overpriced ones.
These strategies are not arbitrage strategies in a textbook sense i.e.
riskless strategies with positive payoff but they are statistical
arbitrage strategies:  they are supposed to give consistently positive
returns in the long run.

The first such test was conducted by \cite{chiras}
who used the implied volatility as a predictor of future price volatility
of the underlying asset. 
More recently Ait-Sahalia {\it et al}
\cite{yared} have proposed  an arbitrage
strategy based on 
non-parametric kernel estimators of the SPD. The idea is the following:
one starts with a diffusion model for the stock price:
\begin{eqnarray}\label{diff}
dS_t &=& S_t ( \mu dt +  \sigma(S_t) dW_t)
\end{eqnarray}
where the instantaneous volatility $\sigma$ is considered to be a deterministic
function of the price level $S_t$.
The function $\sigma(S)$ is then estimated from the historical price series
of the underlying assets using a non-parametric approach.
Under the assumption of a complete market, the state price density may be 
calculated from $\sigma(S)$, yielding an estimator $q^*-{t,T}$.
Another estimator $q_{t,T}$ may 
be obtained by a kernel method as explained above.
If options were priced according to the theory based on the assumption
in Eq.\ref{diff} then one would observe $q=q^*$, a hypothesis which is rejected
by the data. The authors then propose to exploit the difference
between the two distributions to implement a simple trading strategy,
which boils down to buying options for which the theoretical price calculated
with  $q^*$ is lower than the market price (given by $q$) and selling in
the opposite case. They show that their strategy yields a steady profit ($34.5\%$
annualized with a Sharpe ratio around 1.0)
when
tested on historical data. 
Such results have yet to be confirmed on other  markets and data sets
and it should
be noted that large data sets are needed to implement them.

\section{Discussion} \label{summary}

We have described different methods for extracting the statistical information
contained in market prices of options. There are several points which, in our
opinion, should be kept in
mind when using the results of such methods
either in a theoretical context or in applications:

\begin{enumerate}
\item All methods point out to the existence of fat tails,
excess kurtosis
and skewness in the state price density and 
clearly show that the state price density 
is different from a lognormal, as assumed in the Black-Scholes model.
The Black-Scholes formula is simply used as a tool for translating prices into
implied volatilities and not as a pricing method.

\item The study of the evolution of the state price density 
under time aggregation
shows a nontrivial term structure of the implied cumulants, resembling
the terms structure of the historical cumulants.
For example the term structure of the implied kurtosis shows a slow decrease
with maturity which bears a striking similarity with that of historical
kurtosis \cite{adap}.
In the terms used in the mathematical finance literature, the ``risk-neutral"
dynamics is not well described by a random walk / (geometric) 
Brownian motion model.

\item One should {\it not} confuse the state price densities estimated
by the approaches discussed above with the historical densities obtained
from the historical evolution of the underlying asset.
This confusion can be seen in many of the articles cited above:
it amounts to implicitly assuming an expectation pricing rule.
The two densities
reflect two different types of information: while the historical densities
reflects the fluctuations in the market price of the underlying asset,
option prices and therefore the state price density reflects the anticipations
and preferences of market participants rather than the actual (past or future)
evolution of prices. This distinction is clearly emphasized in
\cite{abken} and more explicitly in the maximum entropy method \cite{stutzer}
 where even by minimizing the distance of the SPD with the historical distribution
 one finds two different distributions. Another way of stating this result is
 that option prices are not simply given by historical averages of their payoffs.
 
 \item More specifically, accessing the state price density empirically
 enables a direct comparison with the historical density which provides
 a tool for studying a central question in option pricing theory: the relation
 between the historical and the so-called ``risk-neutral" density.
 The results show that the two distributions
 not only differ in their mean but may also differ in higher moments
 such as skewness or kurtosis.

 In particular the ``intuition" conveyed by the Black-Scholes model
 that the pricing density is simply a centered (zero-mean) version
 of the historical density is not correct. 
 In this sense one sees that
 the Black-Scholes model  is a ``singularity" and its properties should not
be considered as generic.

\item Although this is implicitly assumed by many 
authors, it
is not obvious that the state price densities estimated from options data
do actually correspond to the ``risk-neutral probabilities" or ``martingale measures"
\cite{harrison1,musiela} used in the mathematical
finance literature. 
Although  constraint of the absence of arbitrage opportunities theoretically imposes
that all option prices be expressed as 
expectations of their payoff with respect to the {\it same} density,
the introduction of transaction costs and 
other market imperfections (limited liquidity for example)  can allow for
the simultaneous existence of several SPDs compatible with the observed prices.
Indeed the presence of market imperfections may drastically modify
the conclusions of arbitrage-based pricing theories \cite{figlewski}.
In statistical terms, it is not clear whether a set of  option prices
determine the SPD uniquely in the
presence of market imperfections
{\it even} from a theoretical point of view (e.g. if one
could observe an infinite number of strikes $K$).
This issue has yet to be investigated both from a theoretical and  empirical
point of view.
\end{enumerate}

The methods described above are becoming increasingly common in
applications and will lead to an enhancement of  arbitrage activities between the
spot and option markets. Given the rapid development of  options markets,
 the volume of such
arbitrage trades is not negligible compared to the initial volume of the spot market, giving
rise to non-negligible feedback effects. The existence of feedback implies that
derivative asset such as call options cannot be priced in a framework where
the underlying asset is considered as a totally exogenous stochastic process:
the distinction between underlying and derivative asset becomes less clear-cut
than what text-book definitions tend to  make us think.
The development of such ``integrated" approaches to asset pricing 
should certainly be on the agenda of future research.

I would like to thank Yacine Ait-Sahalia, Jean-Pierre Aguilar, Jean-Philippe Bouchaud, 
Nicole El Karoui,
Jeff Miller and Marc Potters for helpful discussions, {\it Science \& Finance SA}
for their hospitality and the organizers of
the Budapest workshop, Janos Kertesz and Imre Kondor, for their invitation.
The figures are taken from  \cite{adap}.

\end{document}